\documentclass[floatfix,prl,superscriptaddress,nofootinbib,amsmath,amssymb,aps,twocolumn]{revtex4-2}

\usepackage{graphicx}
\usepackage{dcolumn}% Align table columns on decimal point
\usepackage{bm}% bold math
\usepackage{mathrsfs}
\usepackage{xcolor,graphicx}% Include figure files
\usepackage{dcolumn}% Align table columns on decimal point
\usepackage{bm}% bold math
\usepackage{enumerate}
\usepackage{feynmf}
\usepackage{slashed}
\usepackage{subfigure}
\usepackage{todonotes}
\usepackage{multirow}
\usepackage{hyperref}
\usepackage{xspace}
\usepackage{float}
\usepackage{ulem}
\usepackage[T1]{fontenc}
\usepackage[compat=1.1.0]{tikz-feynman}

\newcommand{\gev}{\operatorname{GeV}}
\newcommand{\tev}{\operatorname{TeV}}

\newcommand{\ab}{\operatorname{ab}}

\newcommand{\br}{\operatorname{Br}}

\newcommand{\ie}{\textit{i.e.}}

\newcommand{\customsection}[1]{\vspace{10pt} \noindent \textbf{#1}\vspace{5pt}}

\allowdisplaybreaks[4]

\begin{document}

\title{Distinguishing Dirac from Majorana Heavy Neutrino at Future Lepton Colliders}

\author{Qing-Hong Cao}
\email{qinghongcao@pku.edu.cn}
\affiliation{School of Physics, Peking University, Beijing 100871, China}
\affiliation{Center for High Energy Physics, Peking University, Beijing 100871, China}

\author{Kun Cheng}

\email{kun.cheng@pitt.edu}
\affiliation{PITT PACC, Department of Physics and Astronomy,\\ University of Pittsburgh, 3941 O’Hara St., Pittsburgh, PA 15260, USA}
\affiliation{School of Physics, Peking University, Beijing 100871, China}

\author{Yandong Liu}
\email{ydliu@bnu.edu.cn}
\affiliation{Key Laboratory of Beam Technology of Ministry of Education, School of Physics and Astronomy, Beijing Normal University, Beijing, 100875, China}
\affiliation{Institute of Radiation Technology, Beijing Academy of Science and Technology, Beijing 100875, China}

\begin{abstract}

We propose to identify whether a sterile neutrino is Dirac-type or Majorana-type by counting the peak of the rapidity distribution at lepton colliders. Our method requires only one charged-lepton tagging, and the nature of sterile neutrinos can be pinned down once they are confirmed.

\end{abstract}

\maketitle

\customsection{Introduction}

The existence of antiparticles is a consequence of the theory of quantum mechanics combined with Einstein’s theory of relativity, known under the CPT theorem. Usually, a particle and its antiparticle exhibit opposite charges; however, a particle may be equal to its antiparticle if electrically neutral. Neutrino, hitherto the only charge-neutral fundamental fermion, could be its antiparticle. Such a state of fermion was proposed by Majorana in 1937~\cite{Majorana:1937vz}, which are named Majorana fermions and have not been found yet. It would open a window to new physics beyond the Standard Model (SM) if confirmed.

After the confirmation of massive neutrinos~\cite{Super-Kamiokande:1998kpq,SNO:2001kpb},
the mass origin of the neutrinos and their nature, being Dirac or Majorana fermions, remain unknown. 
The seesaw mechanism offers the most promising avenue to generate neutrino masses~\cite{Minkowski:1977sc,Mohapatra:1979ia,Yanagida:1979as,Gell-Mann:1979vob,Schechter:1980gr,Magg:1980ut,Cheng:1980qt,Lazarides:1980nt,Mohapatra:1980yp,Foot:1988aq,Mohapatra:1986aw,Mohapatra:1986bd} and introduces additional sterile neutrinos (SNs) whose masses range from MeV to the grand unification scale. 
In simple seesaw models, these SNs are Majorana fermions, as demonstrated in various foundational works~\cite{Minkowski:1977sc,Mohapatra:1979ia,Yanagida:1979as,Gell-Mann:1979vob,Schechter:1980gr,Magg:1980ut,Cheng:1980qt,Lazarides:1980nt,Mohapatra:1980yp,Foot:1988aq}. However, in inverse seesaw models~\cite{Mohapatra:1986aw,Mohapatra:1986bd} and UV-complete frameworks such as superstring theories~\cite{Lukas:2000fy} and extra dimension models~\cite{Grossman:1999ra,DeGouvea:2001mz,Rodejohann:2014eka}, these SNs can manifest as Dirac fermions. Searching for SNs and determining whether they are Dirac or Majorana fermions are hot topics in the high energy community, as these findings provide insights into the origin of light neutrino masses and can potentially distinguish the different UV-complete models, such as simplest Type-I seesaw models and others.

The resonant search of SNs has been conducted at the high energy collider including the hadron colliders~\cite{Han:2006ip,Atre:2009rg,Das:2012ze,Dev:2013wba,Arganda:2015ija,Dib:2016wge,Antusch:2016ejd,Degrande:2016aje,Arbelaez:2017zqq,Das:2017nvm,Das:2017gke,Pascoli:2018heg,Abada:2022wvh}, electron hadron colliders~\cite{Das:2018usr,Batell:2022ogj}, electron collider~\cite{Banerjee:2015gca,Han:2022uho,Gu:2022nlj,Gu:2022muc,Mekala:2022cmm} and muon collider~\cite{Chakraborty:2022pcc,Li:2023lkl,Kwok:2023dck,Li:2023tbx,He:2024dwh}.
The Majorana feature of SNs is closely related to lepton number violation (LNV) in new physics (NP). Therefore, the critical point to identifying the nature is counting the lepton number in the final state to determine the existence of LNV. 
The signal of the neutrinoless same-sign di-lepton indicating the LNV can be probed at both low-energy neutrinoless double $\beta$-decay ($0\nu\beta\beta$-decay) experiments~\cite{Schechter:1981bd,BahaBalantekin:2018ppj} and high-energy collider experiments~\cite{Han:2006ip,Atre:2009rg,Dib:2016wge,Das:2017gke,Li:2023lkl}.

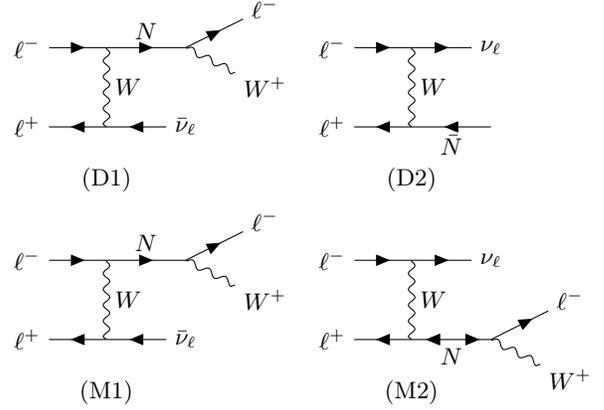
\begin{figure}
    \flushleft
    \begin{tikzpicture}[baseline=(label.base)]
    \begin{feynman}
    \hspace{10pt}
            \vertex (i1) at (-30pt,0pt){$\ell^-$};
            \vertex (i2) at (-30pt,-30pt){$\ell^+$};
            \vertex (v1) at (0pt,0pt);
            \vertex (v2) at (0pt,-30pt);
            \vertex (o1) at (30pt,0pt);
            \vertex (o2) at (30pt,-30pt){$\bar \nu_\ell$};
            \vertex (d1) at (60pt,15pt){$\ell^-$};
            \vertex (d2) at (60pt,-15pt){$W^+$};
            \vertex (label) at (0pt,-50pt){(D1)};
            \diagram[small]{(i1)--[fermion](v1)--[fermion,edge label=$N$](o1)--[fermion](d1),(o1)--[boson](d2),(v1)--[boson,edge label=$W$](v2),(o2)--[fermion](v2)--[fermion](i2)};
        \end{feynman}
    \end{tikzpicture}
    \hspace{10pt}
    \begin{tikzpicture}[baseline=(label.base)]
        \begin{feynman}
            \vertex (i1) at (-30pt,0pt){$\ell^-$};
            \vertex (i2) at (-30pt,-30pt){$\ell^+$};
            \vertex (v1) at (0pt,0pt);
            \vertex (v2) at (0pt,-30pt);
            \vertex (o1) at (30pt,0pt){$\nu_\ell$};
            \vertex (o2) at (30pt,-30pt);
            \vertex (label) at (0pt,-50pt){(D2)};
            \diagram[small]{(i1)--[fermion](v1)--[fermion](o1),(v1)--[boson,edge label=$W$](v2),(o2)--[fermion,edge label=$\bar N$](v2)--[fermion](i2)};
        \end{feynman}
    \end{tikzpicture}
        \begin{tikzpicture}[baseline=(label.base)]
    \hspace{10pt}
        \begin{feynman}
            \vertex (i1) at (-30pt,0pt){$\ell^-$};
            \vertex (i2) at (-30pt,-30pt){$\ell^+$};
            \vertex (v1) at (0pt,0pt);
            \vertex (v2) at (0pt,-30pt);
            \vertex (o1) at (30pt,0pt);
            \vertex (o2) at (30pt,-30pt){$\bar \nu_\ell$};
            \vertex (d1) at (60pt,15pt){$\ell^-$};
            \vertex (d2) at (60pt,-15pt){$W^+$};
            \vertex (label) at (0pt,-50pt){(M1)};
            \diagram[small]{(i1)--[fermion](v1)--[fermion,edge label=$N$](o1)--[fermion](d1),(o1)--[boson](d2),(v1)--[boson,edge label=$W$](v2),(o2)--[fermion](v2)--[fermion](i2)};
        \end{feynman}
    \end{tikzpicture}
    \hspace{10pt}
    \begin{tikzpicture}[baseline=(label.base)]
        \begin{feynman}
            \vertex (i1) at (-30pt,0pt){$\ell^-$};
            \vertex (i2) at (-30pt,-30pt){$\ell^+$};
            \vertex (v1) at (0pt,0pt);
            \vertex (v2) at (0pt,-30pt);
            \vertex (o1) at (30pt,0pt){$\nu_\ell$};
            \vertex (o2) at (30pt,-30pt);
            \vertex (d1) at (60pt,-15pt){$\ell^-$};
            \vertex (d2) at (60pt,-45pt){$W^+$};
            \vertex (label) at (0pt,-50pt){(M2)};
            \diagram[small]{(i1)--[fermion](v1)--[fermion](o1),(v1)--[boson,edge label=$W$](v2),(d1)--[anti fermion](o2)--[anti majorana,edge label=$N$](v2)--[fermion](i2),(o2)--[boson](d2)};
        \end{feynman}
    \end{tikzpicture}
    \caption{Feynman diagrams of the SN production and decay for  (D) Dirac type $N$ and (M) Majorana type $N$.
    Only the decay process with negative charged lepton ($\ell^- W^+$) is plotted. } 
    \label{fig:ll2N2lW_M}
\end{figure}

\begin{figure}
    \centering
    \includegraphics[width=\linewidth]{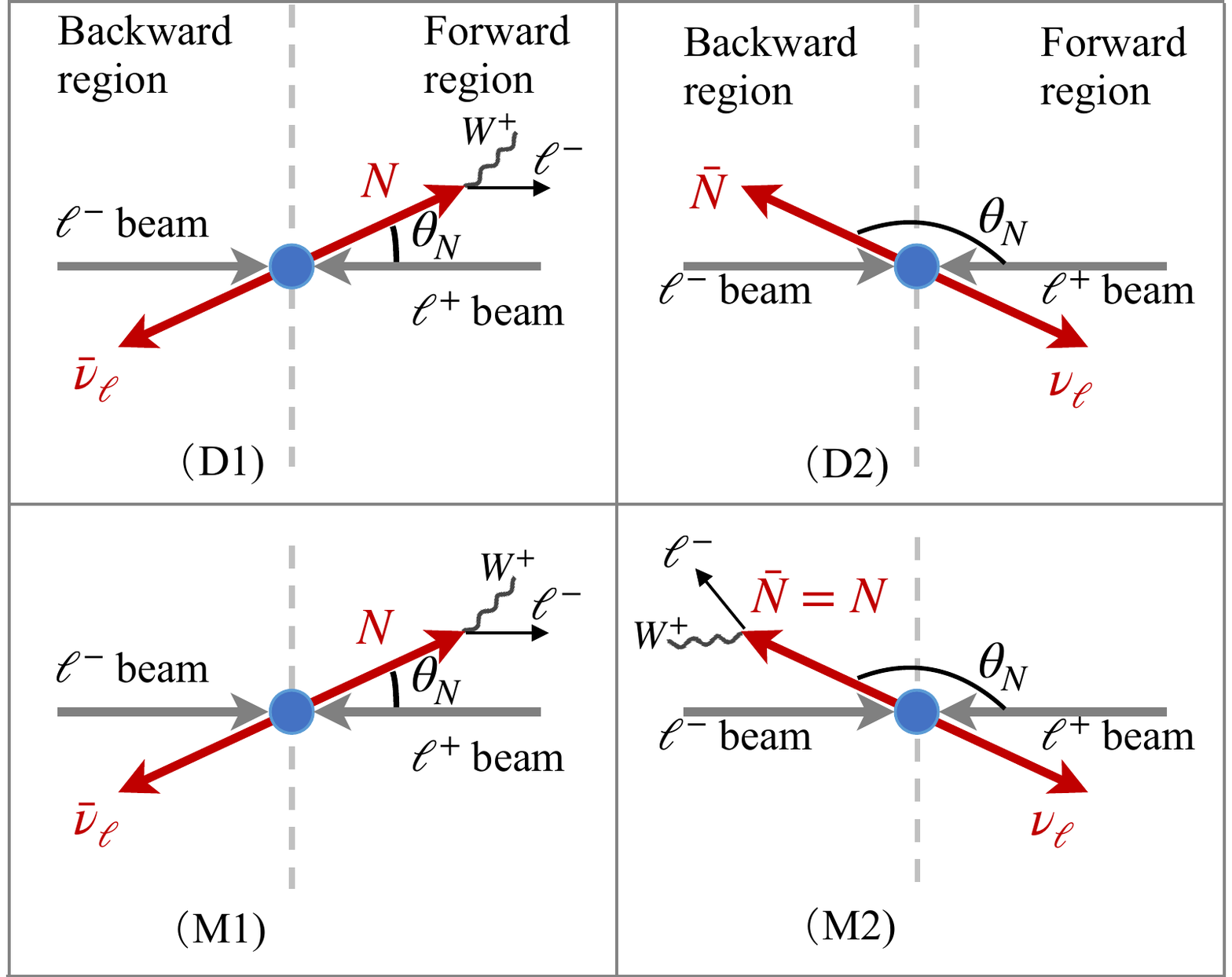}
    \caption{Illustration figure of the SN production.  Dirac-type: only $N$ can decay to $\ell^- W^+$ so  $\ell^-$ are distributed in the forward region.  Majorana-type: $N$ and $\bar N$ are the same particles; therefore, they can decay to $\ell^- W^+$. }
    \label{fig:ll2Nnu}
\end{figure}

In this work, we demonstrate that one can discriminate the nature of SNs in the single SN production at future lepton colliders. 
Consider the $t$-channel resonant production of heavy SNs, \ie, the $\ell^- \ell^+ \to N \bar \nu_{\ell}$ and $\ell^- \ell^+ \to  \nu_{\ell} \bar N$ processes as shown in Fig.~\ref{fig:ll2N2lW_M}. If $N$ is a Dirac-type particle, the two processes produce sterile neutrino $N$ and anti-neutrino $\bar N$, respectively; the $N$ tends to fly along the forward direction (the moving direction of $\ell^-$ beam)
and only decays to $\ell^- W^+$ pair, while $\bar N$ is predominantly produced in the backward region and decays to $\ell^+ W^-$ pair; see Fig.~\ref{fig:ll2Nnu}.  In the Majorana case, the lepton number is no longer conserving as $N=\bar N$, then the SNs produced in the forward and backward regions are the same particle, and they decay to both $\ell^-W^+$ and $\ell^+ W^-$ pairs with the same branching ratio.  Therefore, the resonances of $\ell^-W^+$ in both the forward and backward regions are a clear signature of a Majorana fermion.
More specifically, the sterial neutrinos at lepton colliders are dominantly produced at one or two point in the phase space of $(m_N,y_N)$ for Dirac or Majorana type.
We propose to utilize the numbers of peaks in the rapidity distribution of reconstructed $l^-W^+$ to discrimate the Majorana and Dirac SNs. Compared with lepton-number-counting experiments where at least two same-sign leptons are required, our  method needs only one charged lepton in the final state.

The difference between the kinematics of the Dirac SN and Majorana SN~\cite{Petcov:1984nf} and its decay products~\cite{Balantekin:2018ukw,deGouvea:2021ual,deGouvea:2021rpa} has been explored, especially the charged leptons at lepton colliders~\cite{Hernandez:2018cgc,Mekala:2022cmm,Kwok:2023dck,Mikulenko:2023ezx,Mekala:2023kzo}. 
We demonstrate that analyzing the rapidity distribution of the reconstructed $\ell^- W^+$ offers a clearer distinction between different types of SNs compared to methods that focus on a single decay product, as the single decay product tends to diffuse the distribution, obscuring differences between SN types.
Our findings suggest that it is feasible to promptly identify the nature of SNs at lepton colliders once they are detected.

We aim to highlight the distinct advantages of using the rapidity distributions of sterile neutrinos, in contrast to the scattering angle (or pseudo-rapidity) of leptons decaying from SNs~\cite{Hernandez:2018cgc,Mekala:2022cmm,Kwok:2023dck,Mikulenko:2023ezx,Mekala:2023kzo} or reconstructed SNs~\cite{delAguila:2005pin}, which are essential for distinguishing the nature of SNs and enhancing discovery potential. The rapidity distributions of reconstructed SNs exhibit a sharp peak at $\frac{1}{2}\log s/m_N^2$, even after collider simulation. The position of this peak, uniquely determined by the SN mass, serves as a reliable cross-check for mass measurements obtained from the invariant mass distribution.  

\customsection{Majorana Signal from Rapidity Distribution}

We explore the potential for distinguishing the nature of sterile neutrinos (SNs) 
at high-energy lepton colliders, such as muon colliders, which are particularly sensitive 
to direct searches for SNs. 
We consider the following effective couplings:
\begin{equation}
    \sum_{i=e,\mu,\tau} g U_i \big(\bar{\ell}_{i L} \slashed{W}^- N 
      - \frac{1}{\sqrt{2}c_W} \bar{\nu}_{iL} \slashed{Z} N\big) + \text{h.c.},
\end{equation}
where $g$ is the weak coupling strength, $U_i$ represents the mixing between $N$ 
and the $i$-th generation neutrino, and $N$ generally interacts with all left-handed leptons, 
$\ell_{iL}$.
At high-energy colliders, the $s$-channel process 
$\ell^-\ell^+\to Z^\ast \to N \nu_\ell$ is significantly suppressed~\cite{Li:2023tbx}. 
Consequently, our analysis primarily focuses on the $t$-channel sterile neutrino production 
process, as illustrated in Fig.~\ref{fig:ll2N2lW_M}, while incorporating all relevant 
processes in a detailed simulation. The production rate of $N$ is proportional to $|U_\ell|^2$, while its decay 
branching ratios depend on the specific mixing pattern $U_i$. The differential cross section for the process 
$\ell^-(p_1)\ell^+(p_2)\to N(k_1) \bar{\nu}_\ell(k_2)$ is given by:
\begin{align}
    \frac{{\rm{d}} \sigma}{ {\rm{d}}\cos\theta_N } =&\frac{ g^4 |U_\ell|^2  (1+\cos\theta_N)(s-m_N)^2}{ 64\pi s^2  } \times  \nonumber \\
    \label{eq:dXSdcos}
    &\frac{(s+m_N^2+\cos\theta_N(s-m_N^2)) }{ \large[ (s-m_N^2)(1-\cos\theta_N) +2m_W^2\large]^2 },
\end{align}
where $\theta_N$ is the scattering angle of $N$ relative to the $\ell^-$ beam direction, 
$\sqrt{s}$ is the collision energy, and $m_N$ and $m_W$ are the masses of the sterile neutrino 
and the $W$-boson, respectively. Notably, $N$ tends to propagate along the $\ell^-$ direction.

\begin{figure}
    \centering
    \includegraphics{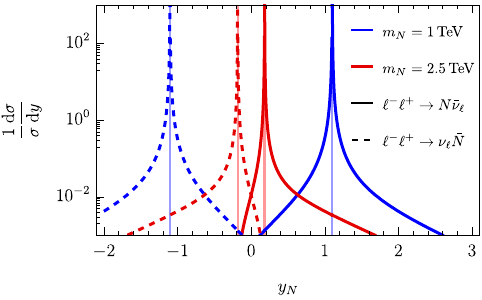}
    \caption{The normalized rapidity distribution of SNs produced from $\ell^-\ell^+\to N \bar \nu_\ell$ (solid lines) and $\ell^-\ell^+\to  \nu_\ell  \bar N$ (dashed lines) with $\sqrt{s}=3\tev$.  The vertical lines are kinematic limit of $y_N$.  Note that $N=\bar N$ for Majorna-type SN. }
    \label{fig:rapidityN}
\end{figure}

The rapidity distribution offers further insights into the collinear production of SNs. 
The rapidity of $N$,
\begin{equation}
    y_N = \frac{1}{2}\log\frac{E+p\cos\theta_N}{E-p\cos\theta_N} \leq \frac{1}{2}\log\frac{s}{m_N^2},
\end{equation}
characterizes the boost of $N$ along the $\ell^-$ direction, where $E$ and $p$ denote 
the energy and momentum of $N$, respectively.
The maximum value of $y_N$, $\frac{1}{2}\log(s/m_N^2)$, follows from the kinematics of the two-body scattering process.
Through a simple algebra the rapidity distribution can be expressed as:
\begin{equation}\label{eq:dXSdy}
    \frac{{\rm{d}} \sigma}{ {\rm{d}}y_N } 
    \propto \frac{e^{4y_N} (s e^{2y_N} - m_N^2)}{(e^{2y_N}+1)^2 
    \big[s + m_W^2 - e^{2y_N}(m_N^2 - m_W^2)\big]^2}.
\end{equation}
This distribution diverges at $\frac{1}{2}\log\big((s + m_W^2)/(m_N^2 - m_W^2)\big)$ 
for $m_N > m_W$ due to the $t$-channel process. When $m_N \ll m_W$, the maximum 
rapidity value, $\frac{1}{2}\log(s/m_N^2)$, is close to the singular point, leading to a sharp peak 
in the rapidity distribution at this value, even after accounting for collider smearing effects 
(see Fig.~\ref{fig:rapidityN}). This behavior aligns with the effective $W$-boson approximation~\cite{Dawson:1984gx, Barger:1990py,Han:2020uid}, 
where the $\ell^-\ell^+ \to N \bar{\nu}_\ell$ process factorizes into two steps: the collinear emission 
of a $W^+$ from the high-energy $\ell^+$ beam, followed by the subprocess $\ell^- W^+ \to N$. 
To produce on-shell $N$, the energy fractions carried by $\ell^-$ and $W^+$ are $x_1 = 1$  and $x_2 = m_N^2/s$ in the massless $W$-boson limit. Consequently, the rapidity of $N$ peaks at $y_p = \frac{1}{2}\log(x_1/x_2) = \frac{1}{2}\log(s/m_N^2)$, consistent with 
Eq.~\eqref{eq:dXSdy} when $m_N \gg m_W$. Take the consequent decay of $N\to \ell^- W^+$ into account, only the forward peak in Fig.~\ref{fig:rapidityN} can be reconstructed from $\ell^{-}W^+$ final states if $N$ is a Dirac fermion, while two peaks are obtained if $N$ is a Majorana fermion.

\customsection{Estimated Sensitivity}

In this section, we explore the potential of the 3 TeV  and 10 TeV muon colliders to discover the SNs and to distinguish the nature of SNs once it is discovered.
We demonstrate the proposed method through a collider simulation on a benchmark case. In the simulation, we demand the gauge boson $W$ hadronic decay as it benefits from large branching ratios and $W$ reconstruction.

\begin{figure}
    \centering
    \includegraphics{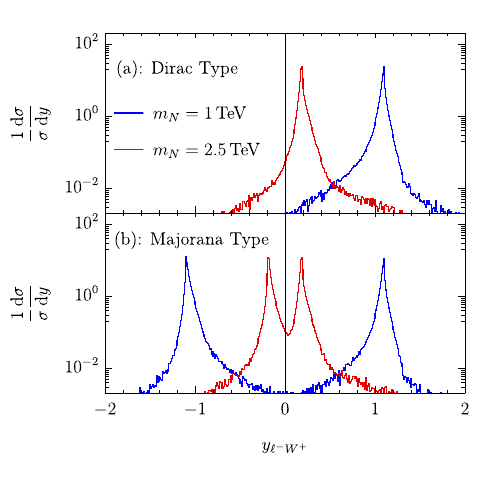}
    \caption{Normalized rapidity distribution of (a) Dirac-type and (b) Majorana-type SN candidates reconstructed from $\ell^-W^+$ pair.  The SNs are produced at $3\tev$ muon collider.}
    \label{fig:delphes-rapidity}
\end{figure}

We generate the events of the $ \mu^{-} \mu^+\to N \bar{\nu}_\mu (\nu_\mu \bar N) $ processes with subsequent decay of $N(\bar N)$ for both Dirac and Majorana SNs with \textsc{MadEvents}~\cite{madgraph5}.  We then pass the events to \textsc{Pythia8}~\cite{pythia} for showering and hadronization and then simulate the collider effects with \textsc{Delphes3}~\cite{delphes}  using the standard CMS card. 
Selected narrow-cone (wide-cone) jets are clustered using the anti-$k_T$ algorithm with radius parameter $R=0.5$ ($R=0.8$)  using \textsc{FastJet}~\cite{fastjet}.  For both lepton and jets, the transverse momentum $p_T$ is required to satisfy $p_T>20\gev$, and the pseudo-rapidity $\eta$ satisfies $|\eta|<2.5$.  Events are required to have one $W$-candidate, one isolated negative-charged lepton, and no additional leptons with $p_T>20\gev$ and $|\eta|<2.5$.
The SNs are reconstructed from a negatively charged lepton and a $W$-candidate, where the $W$-candidate is reconstructed in one of the following two approaches: a) a wide-cone jet $J$ with $50\gev<m_J<100\gev$; b) a pair of narrow-cone jets $j_1 j_2$ with $50\gev<m_{j_1 j_2}<100\gev$ of which the pair with their invariant mass closest to $m_W$ is chosen if several jet pairs are satisfying the cut.  If both reconstruction approaches are available for a single event, preference is given to the $W$-boson reconstructed from a wide-cone jet. 
The rapidity distributions of the reconstructed $\ell^- W$ pairs at a 3 TeV muon collider  are presented in Fig.~\ref{fig:delphes-rapidity}.
For Majorana SNs,  it is observed that the two peaks in the rapidity distribution remain  distinguishable even after accounting for showering and detector effects, as shown in  Fig.~\ref{fig:delphes-rapidity}(b). In contrast, for Dirac-type SNs, only a single peak is  present in the rapidity distribution of $y_{\ell^- W^+}$, specifically in the forward region,  as depicted in Fig.~\ref{fig:delphes-rapidity}(a). Therefore, counting the event number of $\ell^-W^+$ pair produced in forward  and backward regions can distinguish the type of SNs.

The SM backgrounds for SN production vary with the decay mode, e.g., $N \to e^- W^+$ versus 
$N \to \mu^- W^+$, due to beam-induced effects. For simplicity and without loss of generality, 
we focus on the $N \to e^- W^+$ decay mode for illustration, then the dominant SM backgrounds to the signal are $\mu^-\mu^+\to W^+ e^- \bar\nu_e$, $\gamma^\ast\gamma^\ast \to W^+ e^- \bar\nu_e$, $\gamma^\ast \mu^+ \to \bar{\nu}_\mu W^+ e^+ e^-$ and $\mu^- \mu^+ \to \nu_\mu \bar{\nu}_\mu W^+ \bar\nu_e e^-$ processes with hadronic decay $W^+$~\cite{Li:2023tbx}.  
To reject the backgrounds, we require that the invariant mass of the electron and the missing momentum differs from the invariant mass of $W$-candidate more than $20\gev$.  We also require the invariant mass of reconstructed $e^- W^+$ pair within the mass windows $m_N\pm 5\% m_N$.

\begin{figure}
    \centering
    \includegraphics{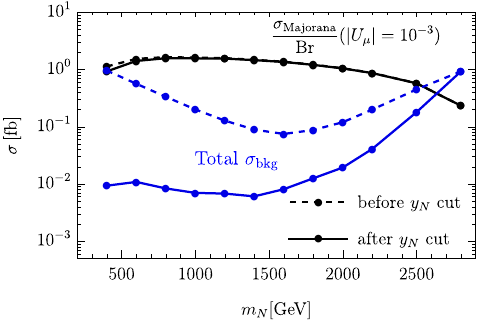}
    \caption{The cross sections of the signal and total backgrounds are shown after applying the selection cuts at the 3 TeV muon collider. 
The dashed lines represent the cases where only the standard selection cuts are applied, 
while the solid lines indicate scenarios where an additional rapidity cut is imposed. }
    \label{fig:effsig}
\end{figure}

 As previously discussed, the rapidity of the 
reconstructed $N$ from the $\ell^-$ and $W$-boson peaks around $\frac{1}{2}\log(s/m_N^2)$, 
a distinctive feature that no Standard Model (SM) background can replicate. 
Consequently, imposing a rapidity cut is expected to significantly reduce background contributions.
In the simulation, following the conventional selection cuts, we further require the reconstructed rapidity 
to lie in a small bin around the peak:
\begin{align} \label{eq:rapiditycut}
y_p-0.1 <|y_{\ell^-W}|< y_p + 0.1, \qquad y_p = \frac{1}{2}\log\frac{s}{m_N^2}
\end{align}
This additional rapidity cut is found to substantially 
suppress the backgrounds with minimal impact on the signal; as illustrated in Fig.~\ref{fig:effsig}.

\begin{figure}
    \centering
    \includegraphics{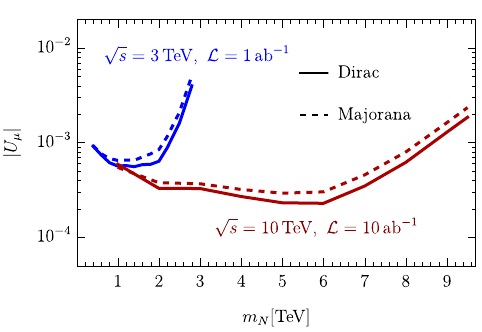}
    \caption{95\% exclusion limit of SN models assuming the SM prediction.}
    \label{fig:exlusionlimit}
\end{figure}

\begin{figure}
    \centering
    \includegraphics{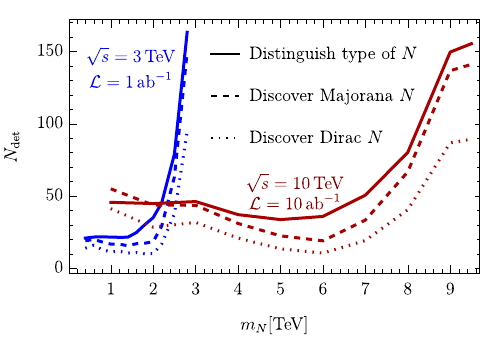}
    \caption{Total number of SN events needs to be detected to claim $5\sigma$ discovery of SNs (dashed line) or to distinguish the SN type (solid line) at $3\tev$ muon collider.  The backgrounds are estimated with an integrated luminosity of $1\ab^{-1}$.  }
    \label{fig:numberofEvents}
\end{figure}

After the above selection cuts, all the events are gathered in two small bins in the phase space of $(M_{\ell^-W^+},y_{\ell^-W^+})$, around $(m_N,y_p)$ and $(m_N,-y_p)$.
To characterize the potential a muon collider to constrain or search for SNs, we first define the likelihood function:
\[
L_{\rm B/D/M} = \prod_{i=1}^2 \frac{\left(n^i_{\rm B/D/M}\right)^{n^i_{\rm obs}} e^{-n^i_{\rm B/D/M}}}{n^i_{\rm obs}!},
\]
where the superscript $i$ denotes the bin index with $y_{\ell^-W^+}>0$ or $y_{\ell^-W^+}<0$, and $n_{\rm B/D/M}$ represents the expected number of events from SM backgrounds ($B$) and signal events predicted by Dirac ($D$) or Majorana ($M$) sterile neutrinos plus the SM backgrounds. The number of SNs $N_{\rm det}$ can easily be translated to parameter spaces as the production cross section is proportional to $|U_\ell|^2$.  We have
\begin{equation}
N_{\rm det}=\mathcal{L}\times\br\times \epsilon_{\rm sel}\times\sigma_{\mu^-\mu^+\to N\bar \nu_\mu},
\end{equation}
where $\mathcal{L}$ is the luminosity, the branching ratio $\br= \br_{W\to jj}\times |U_e|^2/\sum_i |U_i|^2$, and $\sigma_{\mu^-\mu^+\to N\bar \nu_\mu}$ is the signal cross section after selection. For illustration we take ${\rm Br}=\mathrm{Br}_{W\to jj}/2$.
Assuming the SM prediction ($n_{\rm obs}=n_B$), the confidence level for excluding SN predictions is given by
\begin{equation}
    \Delta\chi^2 = -2\log \frac{L_{D/M}}{L_{B}}=2^2.
\end{equation}
We show the 95\% exclusion limit with $\Delta\chi=2$ in Fig.~\ref{fig:exlusionlimit}.
It is found that the coupling can be constrained to around $10^{-3}$ in for SN masses within the reach of c.m. energy.
The exclusion potential (also the discovery potential in Fig.~\ref{fig:numberofEvents}) of Dirac and Majorana type SNs are different because the signal of Majorana SNs are distributed more similar to the SM backgrounds.
In the mass region of our interest, the SN is a narrow resonant that decays promptly after the production, even for a small coupling $|U_\mu|^2\sim 10^{-7}$, while the limit of high-luminosity large hadron collider is $|U_\mu|^2\sim 10^{-3}$~\cite{Li:2023tbx}.

 Apart from the exclusion limit, it is also informative to present the number of event need to reach $5\sigma$ discovery of SNs, and to distinguish the types of SNs.
Assuming the observation corresponds to the prediction from Dirac/Majorana sterile neutrinos 
($n_{\rm obs} = n_{\rm D/M}$), the log-likelihood ratio is given by:
\begin{equation}\label{eq:logL}
    \Delta\chi^2 \equiv -2\log \frac{L_{B}}{L_{\rm D/M}},
\end{equation}
which quantifies the confidence level (C.L.) for discovering sterile neutrino production over SM backgrounds; see the dashed and dotted lines in Fig.~\ref{fig:numberofEvents}.
Similarly, assuming the observation corresponds to the prediction from Majorana sterile neutrinos ($n_{\rm obs} = n_{\rm M}$), 
the log-likelihood ratio
\begin{equation}\label{eq:logL2}
    \Delta\chi^2 \equiv -2\log \frac{L_{\rm D}}{L_{\rm M}},
\end{equation}
quantifies the confidence level (C.L.) for distinguishing Majorana from Dirac scenarios; see the solid lines in Fig.~\ref{fig:numberofEvents}.

At a 3 TeV (10 TeV) muon collider with an integrated luminosity of $1~\mathrm{ab}^{-1}$ ($10~\mathrm{ab}^{-1}$), the number of signal events 
required for discovery is shown in Fig.~\ref{fig:numberofEvents} (dashed and dotted curves). 
Using Eq.~\eqref{eq:logL2}, we calculate the number of signal events required to distinguish Majorana type SNs from Dirac type SNs at the $5\sigma$ confidence level, as shown by the solid curve in Fig.~\ref{fig:numberofEvents}.
As long as the mass of the SNs lies within the energy reach of either the 3 TeV or 10 TeV colliders, only $\sim 10^2$ events are required for both discovery and distinction. The 10 TeV muon collider benefits from a higher collision energy and can probe a wider mass range, while the 3 TeV muon collider has better sensitivity for lower-energy SNs due to a higher detection efficiency.
Remarkably, assuming the predictions of Majorana SNs, the nature of the SN can almost immediately be discerned once they are discovered.  This demonstrates that the kinematic observable, rapidity, effectively captures the LNV signal of the Majorana SNs.

\customsection{Conclusion}

In this work, we propose an approach to distinguishing the nature of SNs, i.e., checking whether the SN is a Majorana or Dirac fermion. 
The SN is tightly related to neutrino mass generation, and its nature will uncover the greatest puzzle in particle physics: whether the Majorana fermion exists. 
It is well recognized that the Majorana SN implies the existence of LNV. Traditionally, the same-sign lepton pair is the key signature to search for the Majorana SNs, including the neutrinoless double beta decay experiments and resonances search at the colliders. However, the lepton number counting always suffers from leaking SM active neutrinos, which carry the lepton number at the colliders.
Taking advantage of the fact that the Dirac or Majorana SNs at lepton colliders are donimation produced at one or two phase space point in the rapidity distribution, our method utilizes only one charged lepton and converts the LNV to counting the peaks of the rapidity distribution of the reconstructed SNs. Only one peak means that the lepton number is conserving,  and the SN is a Dirac fermion; the double peaks mean the lepton number violation, and the SN is a Majorana fermion.

~\\

\begin{acknowledgments}
The work is partly supported by the National Science Foundation of China under Grant Nos. 11725520, 11675002, 11635001, 11805013, 12075257, 12235001.
\end{acknowledgments}

\bibliographystyle{apsrev4-1}
\bibliography{ref}

\end{document}